\documentstyle[preprint,prd,aps,epsf]{revtex}
\draft

\begin{document}
\tightenlines

\title{Effective Gravitational Field of Black Holes}

\author{~Kirill~A.~Kazakov\thanks{E-mail: $kirill@theor.phys.msu.su$}}

\address{Moscow State University, Physics Faculty, Department of
Theoretical Physics, \\ 117234, Moscow, Russian Federation}

\maketitle

\begin{abstract}
The problem of interpretation of the $\hbar^0$-order part of
radiative corrections to the effective gravitational field is considered.
It is shown that variations of the Feynman parameter in gauge conditions
fixing the general covariance are equivalent to spacetime diffeomorphisms.
This result is proved for arbitrary gauge conditions at the one-loop order.
It implies that the gravitational radiative corrections of the order
$\hbar^0$ to the spacetime metric can be physically interpreted in a purely
classical manner. As an example, the effective gravitational field of a
black hole is calculated in the first post-Newtonian approximation,
and the secular precession of a test particle orbit in this field
is determined.
\end{abstract}


\section{Introduction}

The question of interpretation of quantum gravity calculations is
one of the most difficult in quantum field theory. Apart from
difficulties caused by the formal inapplicability of basic notions of
the flat space theory, such as the asymptotic states in the
standard S-matrix approach, the very notion of classical limit in
the quantum theory of gravitation, underlying the issue of
interpretation, is essentially different from that in other
theories of fundamental interactions. It cannot be formulated
neither as the large mass limit of interacting particles, since
the gravitational radiative corrections do not disappear in this
limit \cite{donoghue}, nor even as the formal limit $\hbar \to 0:$
as was shown in Refs.~\cite{kazakov1,kazakov2}, the first
post-Newtonian correction to the gravitational potential, given by
the quantum theory, is twice as large as that given by the
Schwarzschild solution of the classical theory.

It was suggested in Refs.~\cite{kazakov1,kazakov2} that the
correct correspondence between classical and quantum theories is
to be established not for fundamental particles described by field
operators entering the action functional, but rather for {\it
macroscopic} bodies consisting of a large number of such
particles. This interpretation of the correspondence principle is
underlined by an observation that the $n$-loop radiative
contribution to the $n$th post-Newtonian correction to the
gravitational field of a body with mass $M,$ consisting of $N =
M/m$ elementary particles with mass $m,$ contains an extra factor
of $1/N^n$ in comparison with the corresponding tree contribution.
Thus, the effective gravitational field produced by the body turns
into the classical solution of the Einstein equations in the limit
$N \to \infty$ (and therefore, $M \to \infty$). An immediate
consequence of this interpretation is that in the case of {\it
finite} $N,$ the loop corrections of the order $\hbar^0$ describe
deviations of the spacetime metric from classical solutions of the
Einstein equations.

To justify this interpretation completely, one has to prove its
gauge-independence, i.e., that arbitrariness in the choice of gauge
conditions fixing the general covariance does not make values of
measurable quantities, built from the effective metric, ambiguous.
In Ref.~\cite{kazakov1}, independence of the gauge parameter,
weighting the DeWitt gauge conditions in the action, was proved by
direct calculation at the one-loop order. The purpose of this
Letter is to show that this result is not accidental, and to prove
it in a much more simple and general way for arbitrary gauge
conditions. After that, application to the black holes will be
discussed.

\section{Gauge dependence of $\hbar^0$ one-loop corrections}\label{gd}

Let us consider a body with mass $M$ consisting of an arbitrary
number $N$ of particles. For simplicity the latter will be assumed
identical scalars with mass $m,$ denoted by $\phi.$ Dynamics of
the field $\phi$ is described by the action
\begin{eqnarray}&&\label{actionm}
S_{\phi} =  \frac{1}{2}{\displaystyle\int} d^4 x
\sqrt{-g}\left\{g^{\mu\nu}\partial_{\mu}\phi \partial_{\nu}\phi -
\left(\frac{m c}{\hbar}\right)^2\phi^2\right\},
\end{eqnarray}
\noindent while the action for the gravitational
field\footnote{Our notation is $R_{\mu\nu} \equiv
R^{\alpha}_{~\mu\alpha\nu} =
\partial_{\alpha}\Gamma^{\alpha}_{\mu\nu} - \cdot\cdot\cdot,
~R \equiv R_{\mu\nu} g^{\mu\nu}, ~g\equiv \det g_{\mu\nu},
~g_{\mu\nu} = {\rm sgn}(+,-,-,-).$ Dynamical variables of the
gravitational field $h_{\mu\nu} = g_{\mu\nu} - \eta_{\mu\nu},
\eta_{\mu\nu} = {\rm diag}\{+1,-1,-1,-1\}.$}
\begin{eqnarray}&&\label{actionh}
S = - \frac{c^3}{k^2}{\displaystyle\int} d^4 x \sqrt{-g}R,
\end{eqnarray}
where $k^2 = 16\pi G,$ $G$ being the Newton gravitational
constant.

The action $S + S_{\phi}$ is invariant under the following gauge
transformations\footnote{Indices of the functions $F, \xi$ are
raised and lowered, if convenient, with the help of Minkowski
metric $\eta_{\mu\nu}$.}
\begin{eqnarray}\label{gaugesym}
\delta h_{\mu\nu} &=& \xi^{\alpha}\partial_{\alpha}h_{\mu\nu} +
(\eta_{\mu\alpha} + h_{\mu\alpha})\partial_{\nu}\xi^{\alpha} +
(\eta_{\nu\alpha} + h_{\nu\alpha})\partial_{\mu}\xi^{\alpha}
\equiv D_{\mu\nu}^{\alpha}(h)\xi_{\alpha}, \nonumber\\
~~\delta\phi &=& \xi^{\alpha}\partial_{\alpha}\phi \equiv
D^{\alpha}(\phi)\xi_{\alpha},
\end{eqnarray}
where $\xi^{\alpha}$ are the (infinitesimal) gauge functions. Let
this invariance be fixed by the following conditions
\begin{eqnarray}\label{gaugefixnl}
F_{\alpha}(h) = 0.
\end{eqnarray}
\noindent For the beginning, $F_{\alpha}$ will be assumed linear,
\begin{eqnarray}\label{gaugefixlin}
F_{\alpha}(h) \equiv F^{\mu\nu}_{\alpha} h_{\mu\nu},
\end{eqnarray}
\noindent where $F_{\alpha}^{\mu\nu}$ are some differential
operators (Lorentz covariant or not) independent of the fields
$h_{\mu\nu}.$ The most general case will be considered later.
Weighted in the usual way, gauge conditions enter the
Faddeev-Popov action
\begin{eqnarray}\label{fp}
S_{\rm FP} = S + S_{\phi} + S_{\rm gf} +
\bar{C}^{\beta}F_{\beta}^{\mu\nu}D_{\mu\nu}^{\alpha}C_{\alpha}
\end{eqnarray}
\noindent in the form of the gauge fixing term
\begin{eqnarray}\label{gaugefix}&&
S_{\rm gf} = \frac{1}{2\xi} F_{\alpha} F^{\alpha},
\end{eqnarray}
\noindent where $\xi$ is the Feynman gauge parameter, and $C_{\alpha},
\bar{C}^{\alpha}$ are the Faddeev-Popov ghosts.

\begin{figure}\hspace{3cm}\vspace{0,5cm}
\epsfxsize=7cm \epsfbox{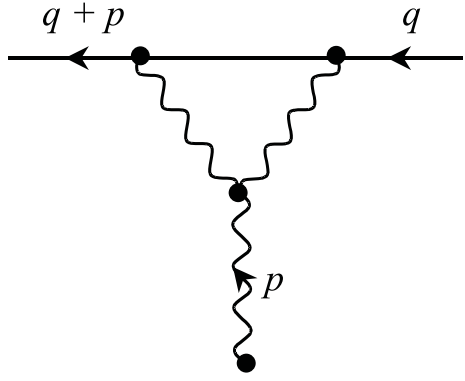} \caption{The one-loop
diagram contributing to the first post-Newtonian correction. Wavy
lines represent gravitons, full lines scalar particles; $q$ and
$p$ are the 4-momentum of the scalar particle and the momentum
transfer, respectively.} \label{fig1}
\end{figure}

Let us now turn to examination of the radiative corrections. Since
we are interested in the quantum contribution to the first
post-Newtonian correction, the only diagram we need to consider is
the one-loop diagram pictured in Fig.~\ref{fig1}. As a simple
dimensional analysis shows, other one-loop diagrams do not contain
root singularities $1/\sqrt{-p^2}$ corresponding to the
$\hbar^0$-contribution, while the higher-loop diagrams are of
higher orders in the Newton constant. Note that it is the
propagation of virtual scalar particle near its mass shell which
is responsible for the occurrence of the $\hbar^0$-contribution.

Although direct calculation of this diagram is cumbersome,
the question of the $\xi$-dependence of its $\hbar^0$
contribution can be easily analyzed as follows.

In the linear gauge (\ref{gaugefixlin}), gauge dependence of this
diagram is determined by that of the graviton propagators. If the
graviton propagator is defined by
\begin{eqnarray}\label{hpr}
\frac{\delta^2 S}{\delta h_{\rho\tau}\delta h_{\mu\nu}}
G_{\mu\nu\zeta\lambda} = - \delta_{\zeta\lambda}^{\rho\tau}\,,\\
~~\delta_{\zeta\lambda}^{\rho\tau} = \frac{1}{2}
(\delta_{\zeta}^{\rho}\delta_{\lambda}^{\tau} +
\delta_{\zeta}^{\tau}\delta_{\lambda}^{\rho})\,, \nonumber
\end{eqnarray}
\noindent then its $\xi$-derivative
\begin{eqnarray}\label{deriv}
\frac{\partial G_{\mu\nu\zeta\lambda}}{\partial\xi} =
G_{\mu\nu\alpha\beta}\frac{F_{\sigma}^{\alpha\beta}
F^{\sigma,\gamma\delta}}{2\xi^2}G_{\gamma\delta\zeta\lambda}\,.
\end{eqnarray}
\noindent On the other hand, multiplying definition (\ref{hpr}) by
the generator $D^{(0)\alpha}_{\rho\tau}\equiv
D^{\alpha}_{\rho\tau}(h=0),$ one has
\begin{eqnarray}\label{slav}
F^{\alpha,\mu\nu} G_{\mu\nu\sigma\lambda} = \xi
D^{(0)\beta}_{\sigma\lambda}\tilde{G}_{\beta}^{\alpha}\,,
\end{eqnarray}
\noindent where $\tilde{G}_{\beta}^{\alpha}$ is the ghost
propagator satisfying
\begin{eqnarray}&&
F_{\alpha}^{\mu\nu}D^{(0)\beta}_{\mu\nu}\tilde{G}^{\gamma}_{\beta}
= - \delta_{\alpha}^{\gamma}\,. \nonumber
\end{eqnarray}
\noindent Let us first consider the inner propagators in
Fig.~\ref{fig1}. In view of Eqs.~(\ref{deriv}), (\ref{slav}),
$\xi$-dependent terms in these propagators are attached to the
scalar line through the generator $D^{(0)\alpha}_{\mu\nu}.$ On the
other hand, the action $S_{\phi}$ is invariant with respect to the
gauge transformations (\ref{gaugesym}),
\begin{eqnarray}&&
\frac{\delta S_{\phi}}{\delta\phi}D^{\alpha}(\phi) + \frac{\delta
S_{\phi}}{\delta h_{\mu\nu}}D^{\alpha}_{\mu\nu}(h) = 0.
\end{eqnarray}
\noindent Differentiating this identity with respect to $\phi,$
setting $h_{\mu\nu} = 0,$ and taking into account that the
external scalar lines are on the mass shell $$\left.\frac{\delta
S_{\phi}}{\delta\phi}\right|_{h=0} = 0,$$ each of the two
$\phi-h-\phi$ vertices can be written as
\begin{eqnarray}&&
\left.\frac{\delta^2 S_{\phi}}{\delta\phi\delta
h_{\mu\nu}}\right|_{h=0}D^{(0)\alpha}_{\mu\nu} = -
\left.\frac{\delta^2
S_{\phi}}{\delta\phi^2}\right|_{h=0}D^{\alpha}(\phi).
\end{eqnarray}
\noindent Thus, under contraction with the vertex factor, the
scalar particle propagator, $G_{\phi},$ satisfying
$$\left.\frac{\delta^2 S_{\phi}}{\delta\phi^2}\right|_{h=0} G_{\phi}
= - 1,$$ cancels out
\begin{eqnarray}\label{cancel}
G_{\phi}\left.\frac{\delta^2 S_{\phi}}{\delta\phi\delta
h_{\mu\nu}}\right|_{h=0}D^{(0)\alpha}_{\mu\nu} = D^{\alpha}(\phi).
\end{eqnarray}
\noindent We conclude that the $\hbar^0$ contribution to the
one-particle-irreducible part of Fig.~\ref{fig1} is
$\xi$-independent. Now, using Eqs.~(\ref{deriv}) and (\ref{slav})
in the external propagator, we see that the $\xi$-derivative of
the corresponding contribution to the whole diagram is
proportional to the generator $D^{(0)\alpha}_{\mu\nu}.$
In other words, variations of the Feynman parameter induce spacetime
diffeomorphisms.

This important result allows us to interpret the $\hbar^0$-part
of the radiative corrections in a purely classical manner.
It was mentioned above that the loop corrections to the spacetime metric
can be endowed with physical meaning only if their gauge dependence
does not introduce an ambiguity into the values of measurable quantities.
The latter are generally defined as functionals of the field variables,
invariant with respect to the spacetime diffeomorphisms.
Since we presently deal with the first post-Newtonian correction, this
criterion can be written
$$\frac{\delta O}{\delta g^{\rm eff}_{\mu\nu}}
D^{\alpha}_{\mu\nu}(h^{\rm eff}) \approx
\frac{\delta O}{\delta g^{\rm eff}_{\mu\nu}}D^{(0)\alpha}_{\mu\nu} = 0,
~~h^{\rm eff}_{\mu\nu} = g^{\rm eff}_{\mu\nu} - \eta_{\mu\nu},$$
where $O$ is any observable, and $g^{\rm eff}_{\mu\nu}$
the effective metric field. We thus see that variations of the Feynman
parameter do not affect values of the observables. This is as it should be,
since, unlike other gauge parameters entering the gauge conditions
$F_{\alpha}$ and determining structure of a given coordinate system,
the weighting parameter $\xi$ does not have any geometrical meaning.
Furthermore, using the $\xi$-independence of observables, one can put
$\xi = 0.$ Then Eq.~(\ref{slav}) shows that the effective metric can
always be chosen to satisfy the gauge conditions exactly,
$$F_{\alpha}(h^{\rm eff}) = F_{\alpha}^{\mu\nu} h^{\rm eff}_{\mu\nu} = 0.$$

Finally, let us consider the most general case of nonlinear gauge
conditions. The gauge fixing term has the form
\begin{eqnarray}\label{nonlin}
S_{\rm gf} = \frac{F^{\alpha}F_{\alpha}}{2\xi} =
\frac{\tilde{F}^{\alpha}\tilde{F}_{\alpha}}{2\xi} +
\frac{\tilde{F}^{\alpha}f_{\alpha}}{\xi} +
\frac{f^{\alpha}f_{\alpha}}{2\xi},
\end{eqnarray}
where $\tilde{F}_{\alpha}$ is the linear part of $F_{\alpha},$
having the form (\ref{gaugefixlin}), and $f_{\alpha}$ is of higher
orders in the fields $h_{\mu\nu},$ responsible for the appearance
of new, ``fictitious'' interactions of gravitons. At the one-loop
order, there is only one new diagram, of the type shown in
Fig.~\ref{fig1}, in which the triple graviton vertex is generated
by the second term in the right hand side of Eq.~(\ref{nonlin}). It
is easy to see that the effect of addition of this diagram is again
a spacetime diffeomorphism. Indeed, if the factor
$\tilde{F}_{\alpha}$ acts on one of the internal graviton lines,
using Eq.~(\ref{slav}) and repeating literally the reasoning which
led to Eq.~(\ref{cancel}), we see that the $\hbar^0$-terms fall
out of the diagram. If the factor $\tilde{F}_{\alpha}$ acts on the
external graviton line, the use of Eq.~(\ref{slav}) shows that the
corresponding contribution is proportional to
$D^{(0)\alpha}_{\mu\nu}.$

Thus, variations of the Feynman parameter are proved to be equivalent
to spacetime diffeomorphisms for the most general gauge conditions.

\section{Effective gravitational field of black holes}

It was mentioned in the Introduction that in comparison with the
classical general relativity, the one-loop contribution to the
first post-Newtonian correction to the gravitational field of a
macroscopic body consisting of $N$ particles is suppressed by the
factor $1/N.$ For instance, if the gauge condition is that of
DeWitt
\begin{eqnarray}\label{dewitt}&&
F_{\alpha} = \partial^{\mu} h_{\mu\alpha} -
\frac{1}{2}\partial_{\alpha} h, ~~h \equiv \eta^{\mu\nu}
h_{\mu\nu},
\end{eqnarray}
\noindent the gravitational potential of a spherically symmetric
body with mass $M$ is \cite{kazakov1}
\begin{eqnarray}&&\label{exampl}
\Phi(r) =  - \frac{G M}{r} + \frac{G^2 M^2}{2 c^2 r^2} - \frac{G^2
M^2}{N c^2 r^2}\,.
\end{eqnarray}
\noindent This suppression of the quantum contribution guaranties
that the classical predictions of general relativity are in
agreement with observations of motion of macroscopic bodies (for
the solar gravitational field, for instance, $1/N \approx m_{\rm
proton}/M_{\odot} \approx 10^{-57}$).

As was explained in Ref.~\cite{kazakov1}, gravitational interaction
of the constituent particles is taken into account in Eq.~(\ref{exampl}),
up to terms of higher order in $G,$ by identifying $M$ as the gravitational
mass of the body. This is legitimate only if the interaction is not
too strong, namely, if its expansion in powers of $G$ is justified.

Let us now consider a situation when evolution of the system of
particles ends up with formation of the horizon. Then the
above condition on the strength of particle interaction inevitably
breaks down at some stage. In the absence of self-consistent
quantum theory of gravitation, nothing can be said about the
ultimate fate of the collapsing matter. What can be said, however,
is that from the point of view of external observer, the number
$N$ is now irrelevant to the gravitational field of the collapsar
(this is a consequence of the ``no hair'' theorem). Made by the
infinite gravitational force indivisible, this object can be
considered as a ``particle''. I will assume that it can be
described by a scalar field with mass $M$ equal to the
gravitational mass of the black hole. Then the one-loop
contribution of the order $\hbar^0$ to the gravitational field of
the black hole, represented in Fig.~\ref{fig1}, is \cite{kazakov1}
\begin{eqnarray}&&\label{correctionp}
h^{{\rm loop}}_{\mu\nu}(p) = - \frac{\pi^2 G^2}{c^2 \sqrt{-p^2}}
\left(3 M^2 \eta_{\mu\nu} + \frac{q_{\mu} q_{\nu}}{c^2}
+ 7 M^2\frac{p_{\mu} p_{\nu}}{p^2}\right).
\end{eqnarray}
\noindent Written down in the coordinate space with the help of
the formulae
\begin{eqnarray}
\int \frac{d^3{\bf p}}{(2\pi)^3}\frac{e^{i{\bf p x}}}{|{\bf p}|}
&=&
\frac{1}{2\pi^2 r^2}\,,\nonumber\\
\int \frac{d^3{\bf p}}{(2\pi)^3}\frac{p_i p_k}{|{\bf
p}|^3}e^{i{\bf p x}} &=& \frac{1}{2\pi^2 r^2}\left(\delta_{i k} -
\frac{2 x_i x_k }{r^2}\right), ~~r^2 \equiv \delta_{i k }x^i x^k
\,,\nonumber
\end{eqnarray}
\noindent equation (\ref{correctionp}) gives, in the static case,
\begin{eqnarray}&&\label{correctionr}
h^{\rm loop}_{00} = - \frac{2 G^2 M^2}{c^2 r^2}\,, ~~h^{{\rm
loop}}_{i k} = \frac{G^2 M^2}{c^2 r^2}\left(- 2 \delta_{i k} +
\frac{7 x_i x_k}{r^2} \right).
\end{eqnarray}
\noindent In order to find complete expression for the metric
in the first post-Newtonian approximation, one has to add the tree
contribution, given by the Schwarzschild solution transformed to
the DeWitt gauge condition (\ref{dewitt}) under which
Eqs.~(\ref{correctionr}) were derived. Using Eq.~(4) of
Ref.~\cite{kazakov1}, we thus obtain the following expression for
the interval
\begin{eqnarray}\label{eff}&&
ds^2 \equiv g^{\rm eff}_{\mu\nu} d x^{\mu} d x^{\nu} = \left(1 -
\frac{r_g}{r}\right) c^2 d t^2 - \left(1 + \frac{r_g}{r} - \frac{7
r^2_g}{4 r^2}\right) d r^2 \nonumber\\&& - r^2 \left(1 +
\frac{r_g}{r} + \frac{7 r^2_g}{4 r^2}\right) (d\theta^2 +
\sin^2\theta\ d\varphi^2),
\end{eqnarray}\noindent where $\theta, \phi$ are the standard
spherical angles, and $r_g = 2 G M/c^2.$

As an application of the obtained result, let us consider one of
the classic effects of general relativity, the orbit precession in
the gravitational field of a spherically symmetric body. Let a
test particle with mass $m$ move in the equatorial plane ($\theta =
\pi/2$) around black hole. Denoting $S_{\rm b}$ the action of the
body, we write the Hamilton-Jacobi equation
$$g^{\mu\nu}_{\rm eff}\frac{\partial S_{\rm b}}{\partial x^{\mu}}
\frac{\partial S_{\rm b}}{\partial x^{\nu}} - m^2 c^2 = 0,$$ where
$g^{\mu\nu}_{\rm eff}$ is the inverse of $g_{\mu\nu}^{\rm eff}.$ A
simple calculation gives, to the leading order,
\begin{eqnarray}\label{actest}
S_{\rm b} &=& - E t + L\varphi \nonumber\\ &+& \int dr
\left[\left(\frac{E'^2}{c^2} + 2 m E'\right) +
\frac{r_g}{r}\left(m^2 c^2 + 4 m E'\right) -
\frac{1}{r^2}\left(L^2 - 2 r_g^2 m^2 c^2\right)\right]^{1/2},
\end{eqnarray} \noindent where $E, L$ are the energy and angular
momentum of the particle, respectively, and $E' = E - m c^2$ its
non-relativistic energy. The first two terms in the integrand in
Eq.~(\ref{actest}) coincide with the corresponding terms of
classical theory, while the third does not, leading to the angular
shift of the perihelion
\begin{eqnarray}\label{prec}
\delta \varphi = \frac{8\pi G M}{c^2 a (1 - e^2)}
\end{eqnarray} \noindent
per period ($a$ and $e$ are the major semiaxis and the
eccentricity of the orbit, respectively), which is to be compared
with the classic result
$$\delta \varphi = \frac{6\pi G M}{c^2 a (1 - e^2)}\ .$$

\section{Conclusions}

Interpretation of the correspondence principle, suggested in
Ref.~\cite{kazakov1}, endows the $\hbar^0$ loop contributions with direct
physical meaning as describing deviations of the spacetime metric from
classical solutions of the Einstein equations.
This purely classical treatment of the $\hbar^0$ loop contributions
is supported by the main result of the present work: variations of the
weighting parameter are equivalent to spacetime diffeomorphisms. Thus,
observables built from the effective metric are independent of the
unphysical Feynman parameter. As to other gauge parameters entering the
gauge conditions $F_{\alpha}$ and determining geometry of a given
coordinate system, the corresponding analysis is much more complicated
and will be presented elsewhere.

It should be clear from the considerations of Sec.~\ref{gd} that this
result is valid whatever matter fields produce the gravitational field.
In essential, it is a consequence of the gauge symmetry of the action
$S + S_{\phi}.$ Despite simplicity of the proof at the one-loop order,
however, the author has not yet been able to extend it to all orders.

As an application of this result, the one-loop effective
gravitational field of black hole was calculated. It is given by
Eq.~(\ref{eff}). One of the consequences is that the orbit
precession in this field differs from that predicted by the
classical Einstein theory, and is given by Eq.~(\ref{prec}). It
should be mentioned in this connection that emission of the
gravitational waves by the black hole binaries also must be
affected by the quantum contributions. The LIGO and VIRGO
gravitational wave detectors, which are currently under
construction, will hopefully bring light into this issue.

\begin{acknowledgements}
I thank P.~I.~Pronin, K.~V.~Stepanyantz, and especially
A.~V.~Borisov (Moscow State University) for interesting
discussions.
\end{acknowledgements}

\end{document}